\documentclass[3p,twocolumn]{elsarticle}

\usepackage{lineno,hyperref}
\usepackage{amsmath} 
\usepackage{graphics} 
\usepackage{subfig} 
\usepackage{amssymb}
\usepackage{multirow}

\journal{Eur. Phys. J. Plus}

\begin{document}

\begin{frontmatter}

\title{Characterization of TRIGA RC-1 neutron irradiation facilities for radiation damage testing}

\author[mib,infnmib]{Davide Chiesa\corref{mycorrespondingauthor}}
\cortext[mycorrespondingauthor]{Corresponding author}
\ead{davide.chiesa@mib.infn.it}

\author[enea]{Mario Carta}
\author[enea]{Valentina Fabrizio}
\author[enea]{Luca Falconi}
\author[enea]{Angelo Grossi}
\author[mib,infnmib]{Massimiliano Nastasi}
\author[enea]{Mario Palomba}
\author[mib,infnmib]{Stefano Pozzi}
\author[mib,infnmib]{Ezio Previtali}
\author[infnmib]{Pier~Giorgio Rancoita}
\author[mib,infnmib]{Barbara Ranghetti}
\author[mib,infnmib]{Mauro Tacconi}

\address[mib]{Dipartimento di Fisica ``G. Occhialini", Universit\`a degli Studi di Milano - Bicocca, 20126 Milano, Italy}
\address[infnmib]{INFN sezione di Milano Bicocca, 20126 Milano, Italy}
\address[enea]{ENEA Centro Ricerche Casaccia, 00123 S. Maria di Galeria (Roma), Italy}

\begin{abstract}
This paper presents the results of neutron flux measurements at two irradiation facilities of the TRIGA Mark II reactor at ENEA Casaccia Research Center, Italy. 
The goal of these measurements is to provide a complete characterization of neutron irradiation facilities for accurate and precise dose evaluation in radiation damage tests and, more generally, for all applications that need a good knowledge of neutron flux in terms of intensity, energy spectrum and spatial distribution.
The neutron activation technique is used to measure the activation rates of several reactions, chosen so to cover the whole energy range of neutron flux spectrum. 
A multi-group neutron flux measurement is obtained through an unfolding algorithm based on a Bayesian statistical model.
The obtained results prove that this experimental method allows to measure the total neutron flux within 2\% statistical uncertainty, and to get at the same time a good description of its energy spectrum and spatial distribution.   
\end{abstract}


\end{frontmatter}


\section{Introduction}

The prediction of electronic device degradation due to radiation allows one both the correct design of experiments in which high radiation dose will be released on instrumentation and the qualification of microelectronics for space missions. Irradiation tests of such devices allow to predict their end of life behavior and performance degradation. The use of fast neutrons, that induce a negligible total ionizing dose, permits to investigate damage effects due to atomic displacements, which are usually the most relevant for semiconductors. Irradiation facilities where tests are performed must be characterized to accurately determine the neutron spectral fluence and, thus, the imparted NIEL (Non-Ionizing Energy Loss) dose~\cite{LeroyRancoita}. Fast neutrons are produced in nuclear reactors and in accelerator-based neutron sources~\cite{frost2009,TRIUMF1,TRIUMF2,ANITA1,ANITA2,LosAlamos,OakRidge}. 
In the latter case, it is difficult to perform an accurate and precise evaluation of the imparted NIEL dose, because at spallation sources the flux spectrum usually includes also charged particles and high energy neutrons with energies $>$20~MeV, the cross sections of which are known with poor accuracy. Conversely, in fission reactors the spectrum of fast neutrons is well-known~\cite{watt} and a more detailed neutron flux characterization can be carried out. Therefore, since decades, the fast neutrons produced in nuclear reactors are those mainly employed for displacements damage investigations regarding microelectronic devices.

\begin{figure*}[t!]
\centering
\subfloat{\includegraphics[width=0.48\textwidth]{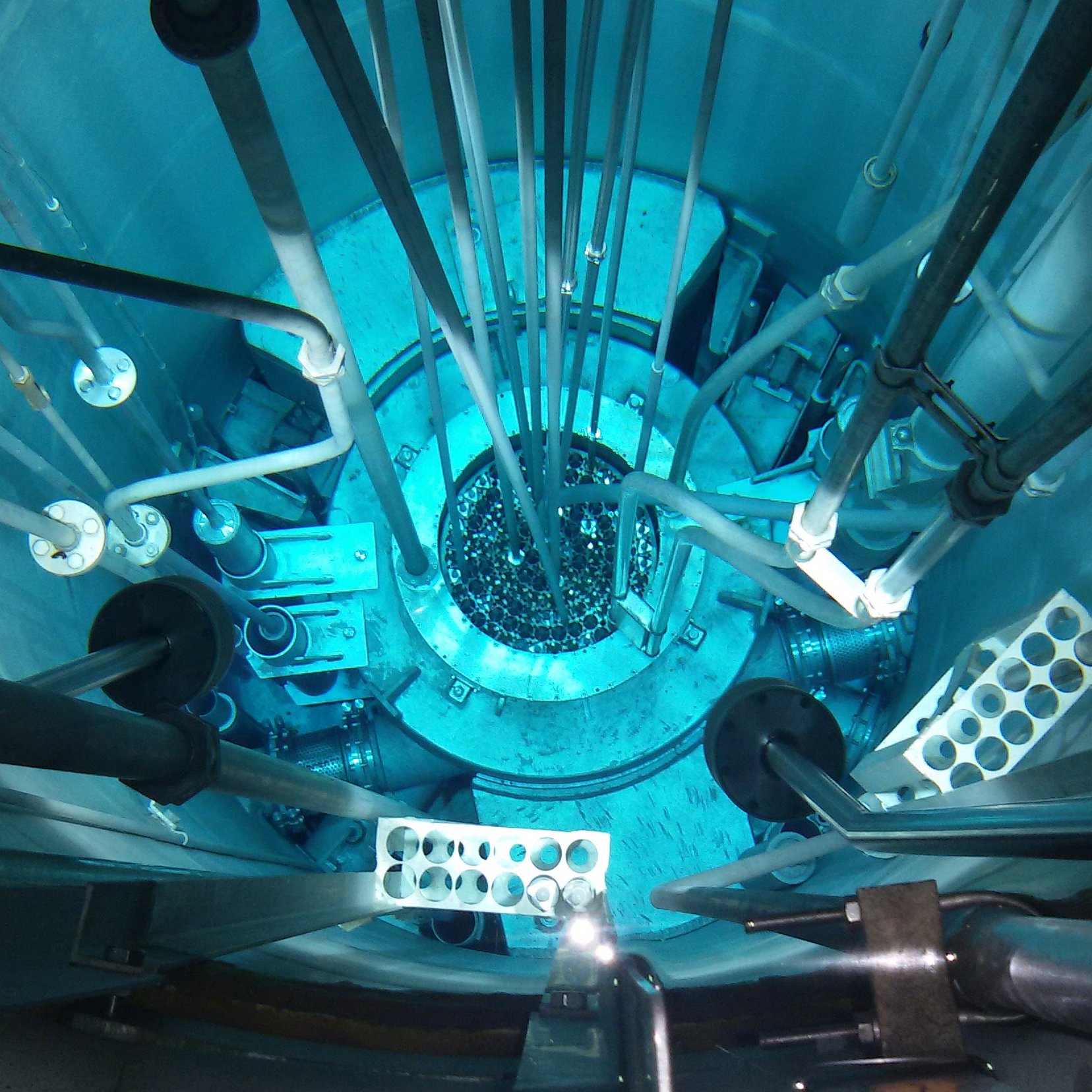}} \hspace{3mm}
\subfloat{\includegraphics[width=0.48\textwidth]{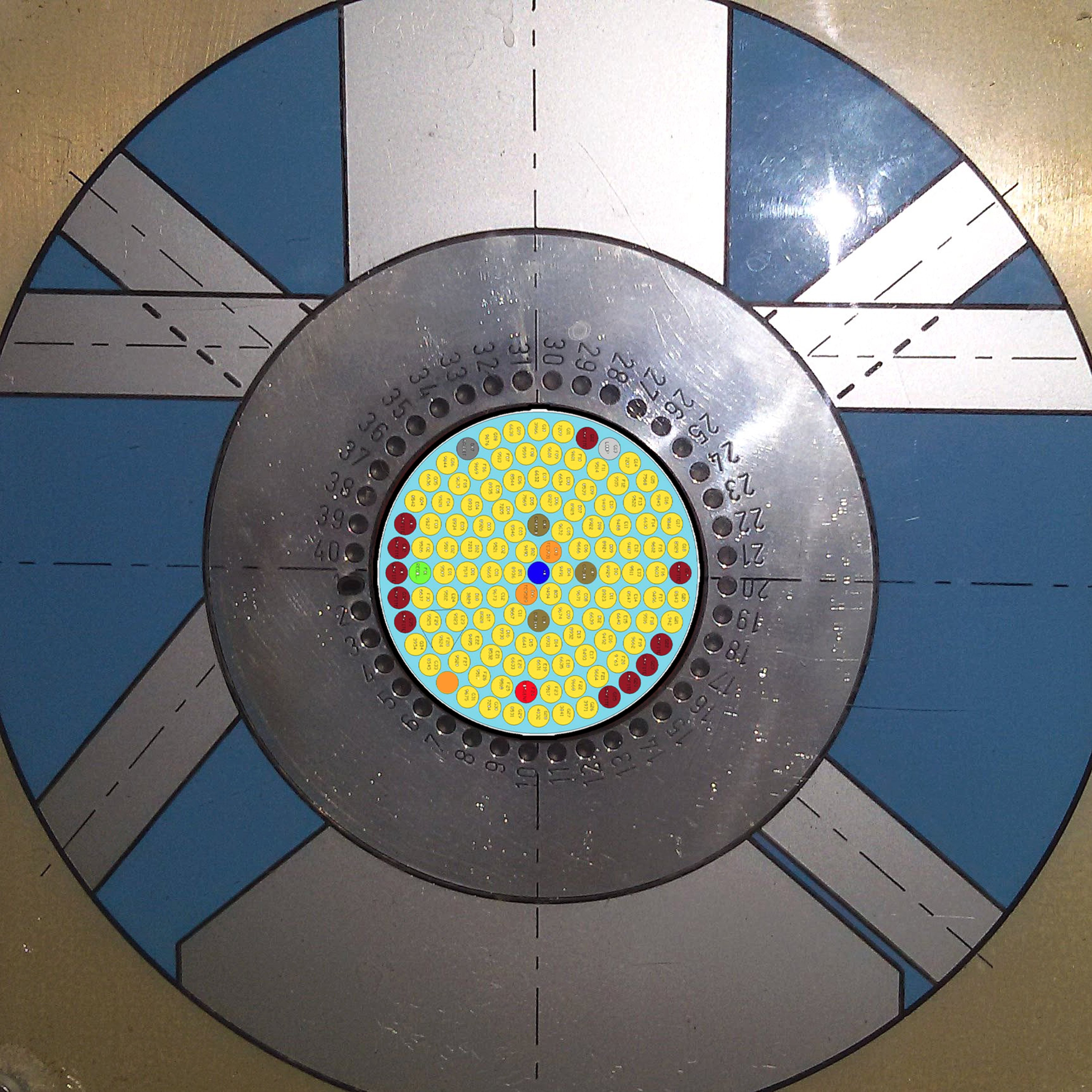}}
\caption{Left: Picture of the TRIGA-RC1 reactor core. Right: Horizontal section of the TRIGA-RC1 reactor core, reflector, and irradiation facilities therein. The core scheme in the center represents the Central Channel (blue), the fuel elements (yellow), the instrumented fuel elements (orange), the \textsc{Shim} control rods (olive green), the \textsc{Regulating} control rod (green), the Am-Be neutron source (red), the graphite \textit{dummy} elements (dark red), the Rabbit Channel (grey), and the experimental loop (light grey). The positions of the 40 channels of Lazy Susan facility are shown around the core.}
\label{Fig:TRIGA}
\end{figure*}

The TRIGA Mark II reactor located at ENEA Casaccia Research Center (TRIGA RC-1) is a research reactor moderated and cooled by light water, and is operated at a maximum thermal power of 1 MW~\cite{ReportCasaccia1969}. The fuel is a uniform mixture of zirconium hydride (ZrH) and uranium (8.5\% weight fraction, 19.9\% enriched in $^{235}$U). The core contains 111 TRIGA standard stainless-steel cladded fuel elements arranged in concentric rings, and is surrounded by a graphite reflector (Fig.~\ref{Fig:TRIGA}). TRIGA RC-1 is equipped with various experimental channels and irradiation positions, providing a wide range of neutron fluxes and spectra useful for several applications~\cite{iaea2017}.
During 2017 an agreement was signed between ENEA, INFN and the Italian Spatial Agency to cooperate in the field of neutron radiation damage analysis on electronic components to be used in future space-crafts. This agreement provides for use ENEA TRIGA RC-1 research reactor as a facility to perform neutron irradiation on such electronic devices. 
This paper describes the experimental characterization of the neutron flux spectrum in the Central Channel and in the Lazy Susan (a specimen rack located around the core) irradiation facilities of the TRIGA RC-1 reactor.

\section{Experimental methods}
Neutron activation is an experimental technique that can be flexibly adapted to characterize different neutron fields~\cite{AbsoluteFlux,ChipIR}, providing an absolute measurement of the neutron flux intensity, energy spectrum~\cite{BayesianSpectrum,ND2013}, and spatial distribution~\cite{FluxDistribution}. 
This method consists of irradiating samples with known amount of elements to produce radioisotopes via neutron-induced reactions.
After the irradiation, $\gamma$-ray spectroscopy measurements are performed to evaluate the activity $A(t)$ of each different radioisotope, which is proportional to the activation rate $R$ of the corresponding reaction:

\begin{equation}
A(t) = R (1-e^{-\lambda t_{\text{irr}}}) e^{-\lambda t}
\label{Eq:Activity}
\end{equation}
where $t_{\text{irr}}$ is the irradiation time and $\lambda$ is the decay constant.
Finally, the experimental measurements of $R$ are used to extract information about the neutron flux $\phi(E)$, exploiting the following physics relation:

\begin{equation}
\label{Eq:ActRate}
R = \mathcal{N} \int dE \, \sigma(E) \phi(E)
\end{equation}
where $\mathcal{N}$ is the number of target isotopes in the sample, and $\sigma(E)$ is the activation cross section.

To determine the neutron flux intensity and its energy spectrum, we apply the unfolding method
described in Ref.~\cite{BayesianSpectrum}. As compared with other unfolding techniques~\cite{Reginatto2010}, the Bayesian approach provides rigorous propagation of experimental uncertainties and allows to include \textit{a priori} information about the neutron flux.
This technique requires measuring the activation rates of several reactions and relies on the fact that the activation cross sections have different energy dependence, thus each reaction is a neutron-sensitive probe in specific energy ranges. 
The reactions are chosen with the criterion of diversifying as much as possible the energy regions of higher neutron-sensitivity, thus improving the accuracy of the unfolded results.
Radiative capture $(n,\gamma)$ reactions are particularly suitable for measuring the neutron flux in the thermal range and in the intermediate one where capture resonances show up, whereas fast neutrons can be effectively measured exploiting threshold reactions such as, for example, $(n,p)$, $(n,\alpha)$, $(n,n')$, $(n,np)$, and $(n,2n)$.

\subsection{Irradiated samples}
\begin{figure*}[!htb]
\centering
\subfloat{\includegraphics[height=4cm]{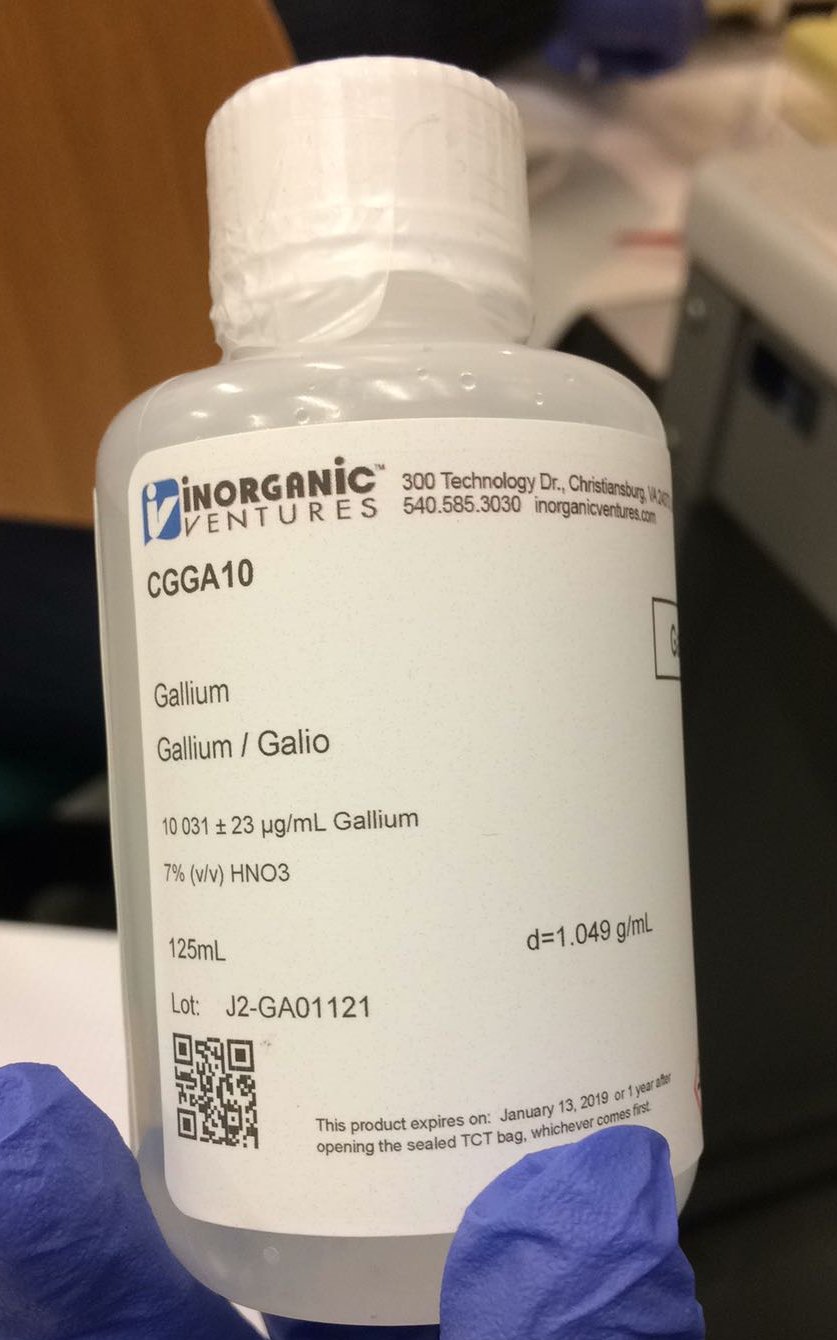}} \hspace{3mm}
\subfloat{\includegraphics[height=4cm]{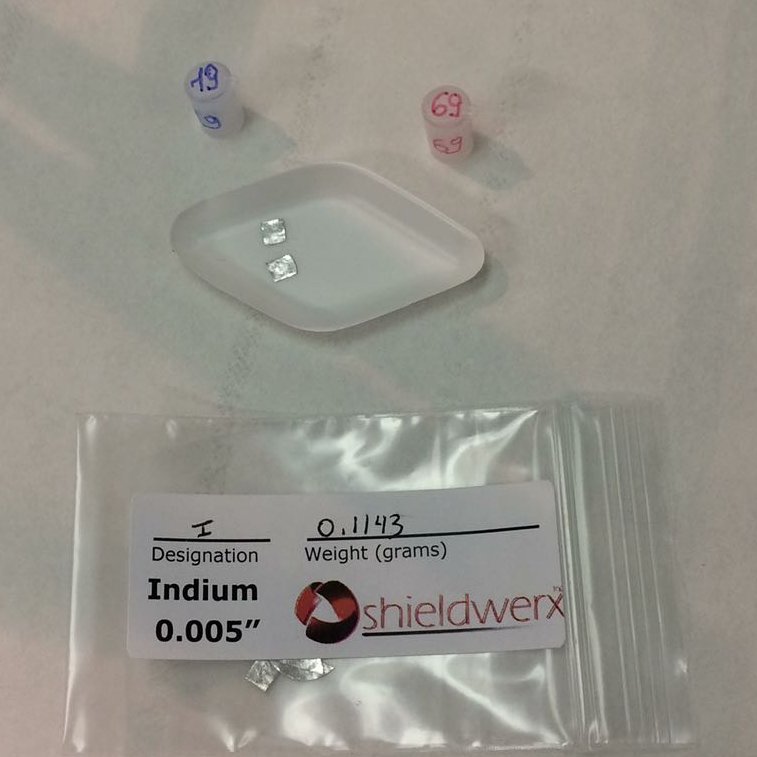}}
\hspace{3mm}
\subfloat{\includegraphics[height=4cm]{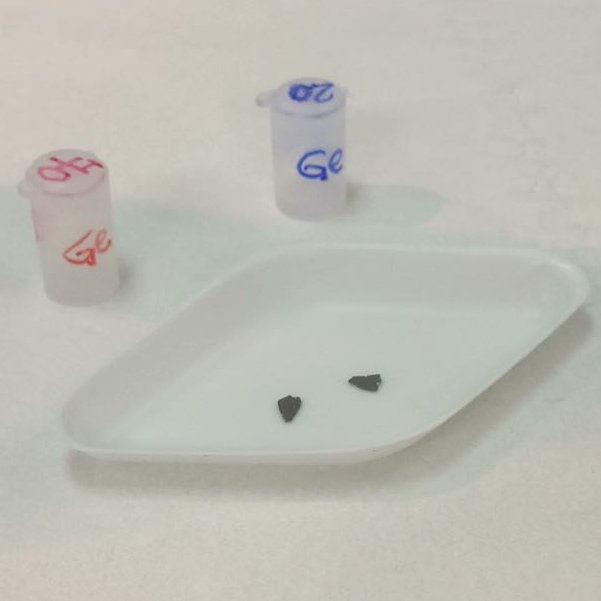}}
\hspace{3mm}
\subfloat{\includegraphics[height=4cm]{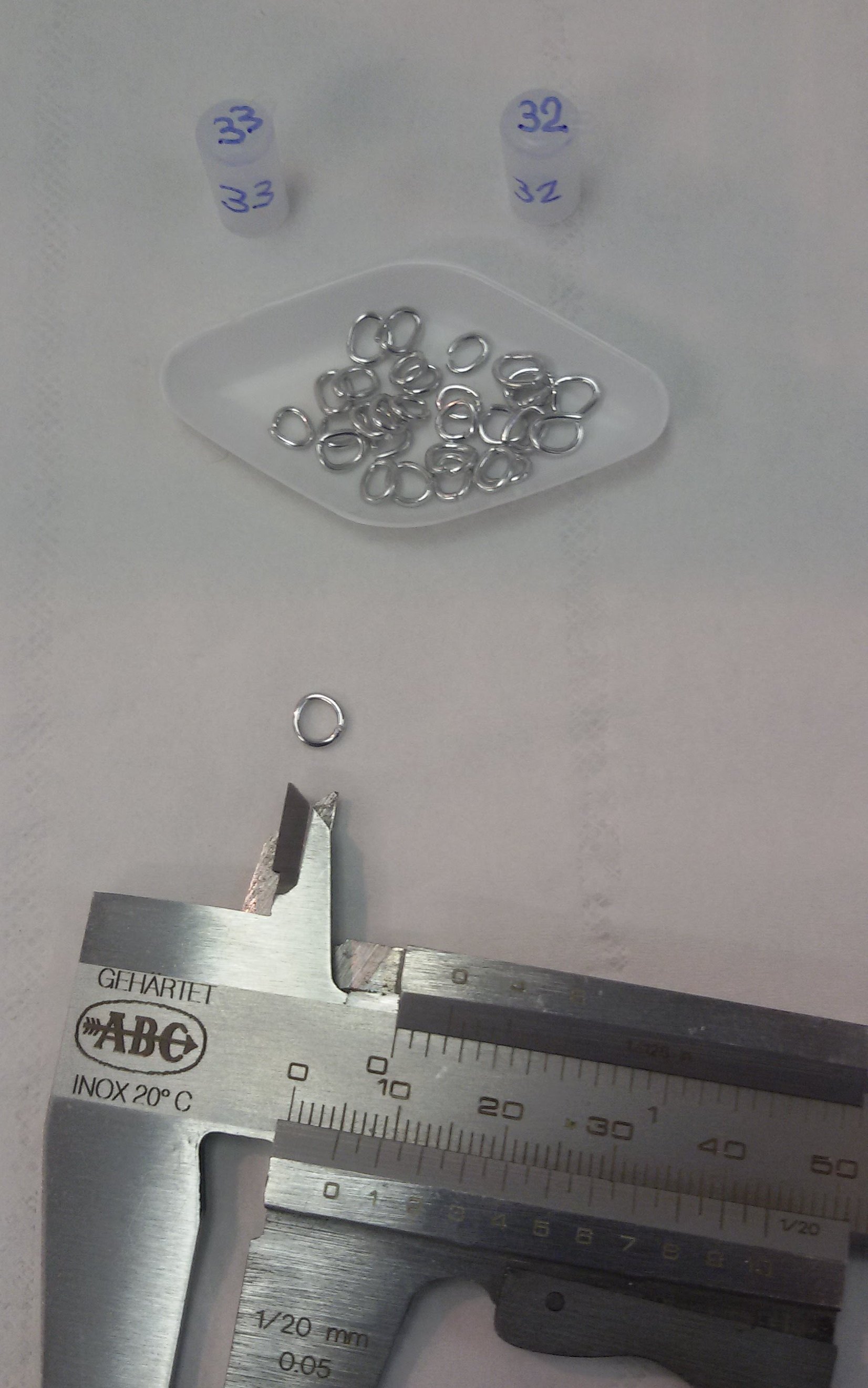}}\\
\subfloat{\includegraphics[height=4cm]{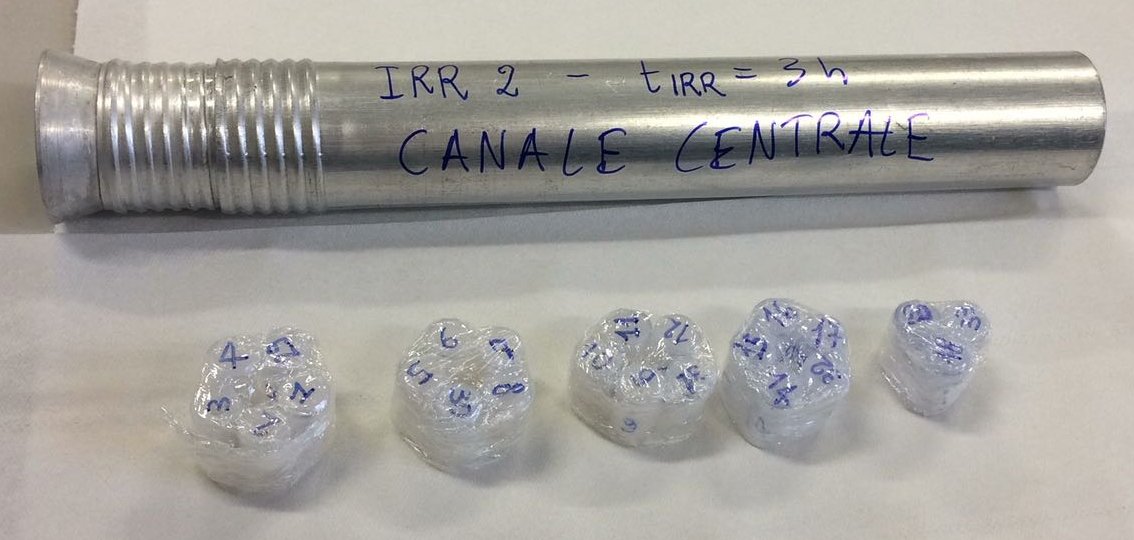}}
\caption{Top: materials used to prepare the samples to be irradiated. From left to right: a liquid standard with certified concentration of trace elements, high purity metal foils cut in small squares to fit into the PET vials, fragments of Ge crystal, Al-Co wires cut and shaped as small rings. \\Bottom: the PET vials packed to fit into the  aluminum holder used for irradiations in the TRIGA-RC1 Central Channel.}
\label{Fig:Samples}
\end{figure*}

To characterize the neutron flux in the Central Thimble and Lazy Susan facilities of the TRIGA-RC1 reactor, we prepared a set of samples containing several elements. 
We used liquid standards with certified concentration of trace elements (ranging from 10 to 10$^4$ $\mu$g/mL), high purity metal foils (0.13~mm or 0.25~mm thick), fragments of ZnSe and Ge crystals (1.8~mm and 0.35~mm thick, respectively), and Al-Co (Co 0.5\% wt.) wires 1~mm in diameter (Fig.~\ref{Fig:Samples}, top). 
Each sample was weighted on a balance with 1~$\mu$g sensitivity, to get precise measurements of the amount of target isotopes. 
The samples were contained in small-sized PET vials (1.5~cm height and 0.8~cm in diameter), that were packed and stacked to fit into the irradiation holders used at the TRIGA-RC1 reactor facilities (Fig.~\ref{Fig:Samples}, bottom). 

\subsection{Neutron irradiations}
The neutron activation experimental campaign has been organized in three separate irradiations, carried out on consecutive days (Tab.~\ref{Tab:Irr}).
In the first one, we irradiated samples containing different elements to allow for the neutron spectrum unfolding analysis. 
The second irradiation served to characterize the neutron flux spatial profile with the technique outlined in Ref.~\cite{FluxDistribution}, based on the activation of several Al-Co samples in different positions of Central Channel and Lazy Susan facilities.
The third irradiation served to activate an indium metal sample (used for the spectrum unfolding analysis) that could not be included in the first irradiation because of the local flux perturbation caused by its high neutron absorption cross section.

\renewcommand{\arraystretch}{1.2}
\begin{table}
\begin{center}
\begin{tabular}{c|c|c}
Irr. \# & Date \& Time & $t_\text{irr}^{\textit{eff}}$ (s) \\
\hline
1	&	03/19/2018 09:16	&	10752	\\
2	&	03/20/2018 10:13	&	3707	\\
3	&	03/21/2018 14:07	&	3743	\\
\end{tabular}
\end{center}
\caption{List of irradiations, with starting dates and effective irradiation times.}
\label{Tab:Irr}
\end{table}

\begin{figure}[!htb]
\centering
\includegraphics[width=0.49\textwidth]{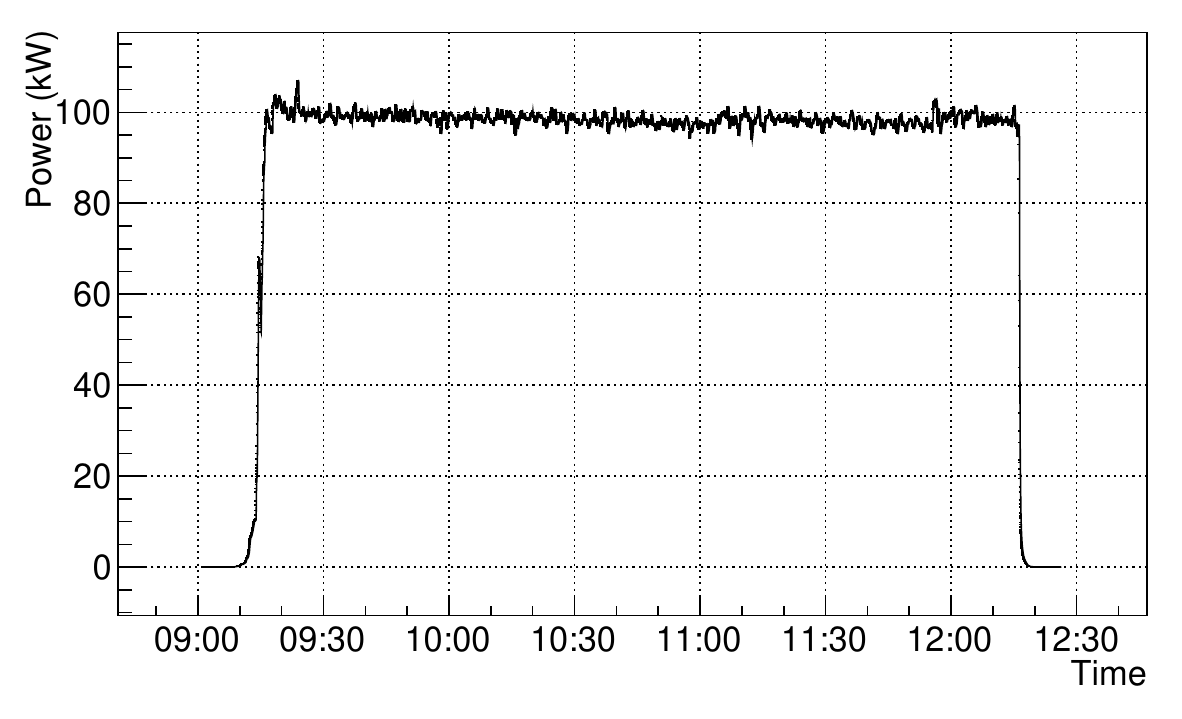}
\caption{Power track of the first irradiation.}
\label{Fig:Irr}
\end{figure}

All irradiations have been performed with the reactor at 100~kW nominal power. We recorded the power tracks (Fig.~\ref{Fig:Irr}) to allow for a precise evaluation of the neutron fluence and of the \textit{effective} irradiation time ($t_\text{irr}^{\textit{eff}}$ calculated as $\int P(t)\,dt / \text{100~kW}$) to keep into account the rise and drop irradiation transients.
During irradiations, we also monitored the reactor parameters that could affect the neutron flux: fuel and water temperature, and control rods positions. These parameters, at 100~kW power, exhibit relatively small excursions, thus we consider the irradiation conditions to be stable.
To have a repeatability test of activation measurements in the different irradiations, we used Al-Co samples as neutron flux monitors in specific positions of Central Channel and Lazy Susan facilities.

\begin{figure*}[!htb]
\centering
\subfloat{\includegraphics[width=0.45\textwidth]{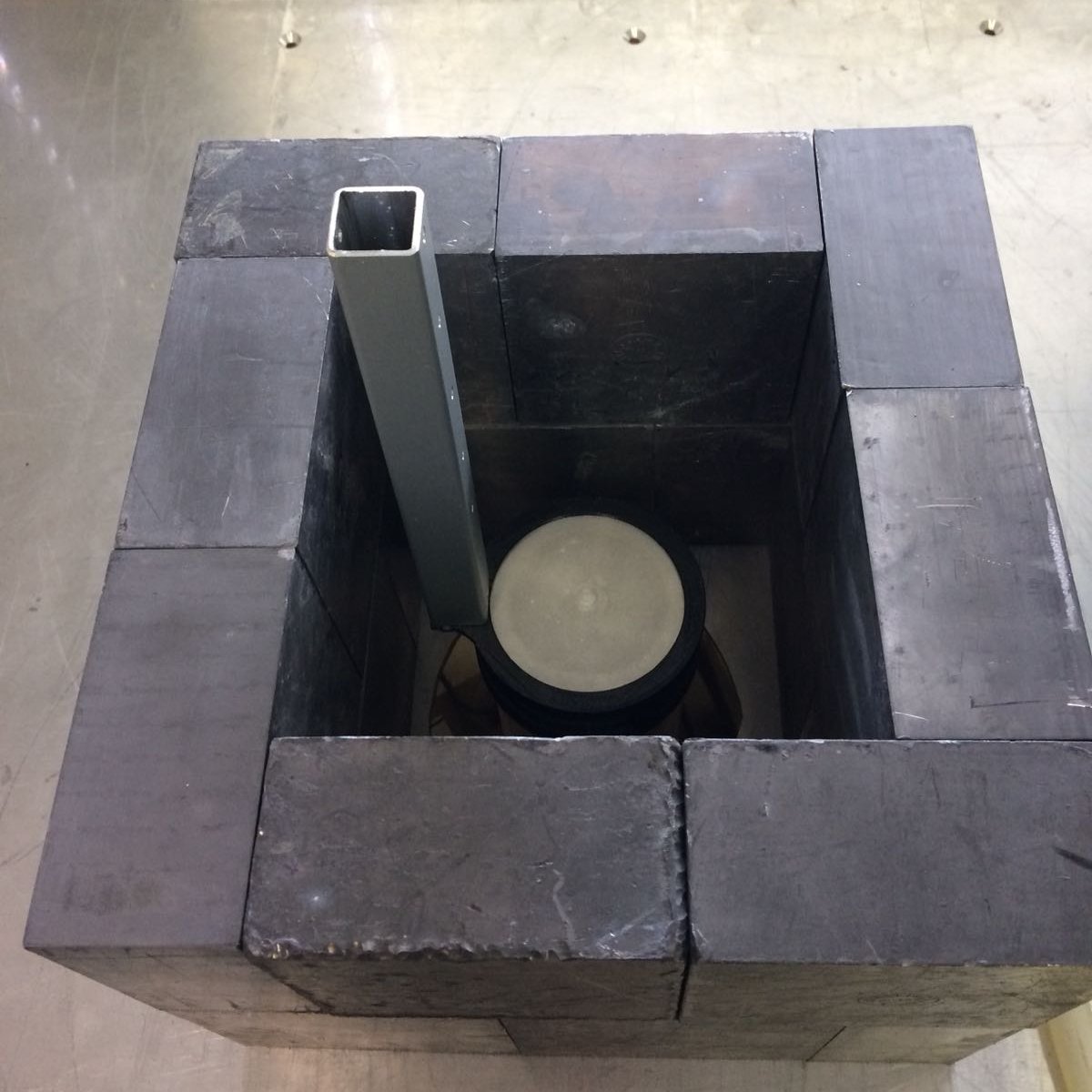}} \hspace{1cm}
\subfloat{\includegraphics[width=0.45\textwidth]{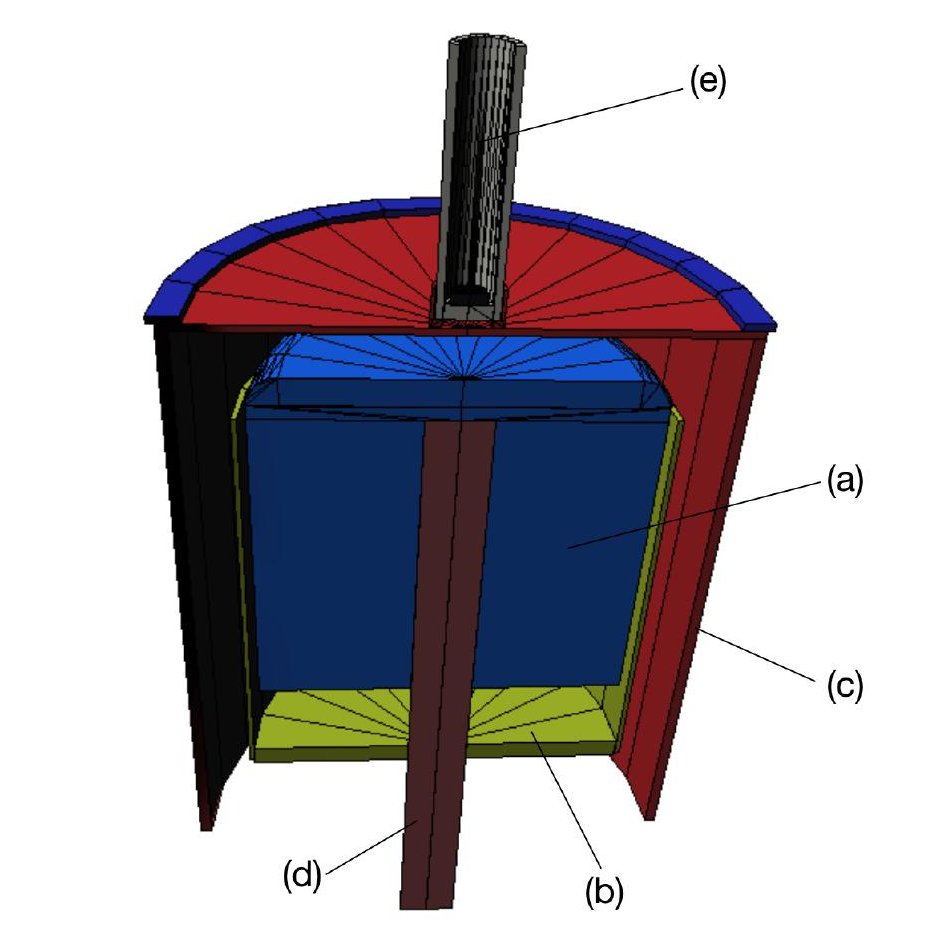}}
\caption{Left: top view of the Silena HPGe detector endcap surrounded by a lead shield and equipped with an holder for sample placement at different distances. Right: geometry of the Silena HPGe detector reconstructed with the \textit{Arby} simulation toolkit: (a) Ge crystal, (b) Cu holder, (c) aluminum endcap, (d) cold finger, (e) sample in a PET vial.}
\label{Fig:HPGe}
\end{figure*}

\subsection{$\gamma$-spectroscopy measurements}

To measure the activity of the radioisotopes produced in the irradiated samples, we set up two High-Purity Germanium (HPGe) detectors (coaxial, p-type) at the ENEA-Casaccia laboratory. 
The first one is a Silena detector with 30\% relative detection efficiency and 1.8~keV FWHM resolution at the 1.33~MeV gamma line of $^{60}$Co (Fig.~\ref{Fig:HPGe}, left).
The second one is a Canberra detector with 42\% relative efficiency and 1.9~keV FWHM resolution at 1.33~MeV.
Both detectors were equipped with a holder structure specifically designed to allow for a precise and repeatable sample placement at various distances from the detector. 

The radioisotope activity is measured from the number of counts in their $\gamma$-lines, with the method described in Refs.~\cite{AbsoluteFlux,ChipIR}. 
This method makes use of Monte Carlo (MC) simulations to get accurate assessment of detection efficiency in any sample-detector geometrical configuration.
For this purpose, we developed simulation models for both HPGe detectors using the \textit{Arby} toolkit, that allows to flexibly implement the sample-detector geometry (Fig.~\ref{Fig:HPGe}, right) and run a \textsc{Geant4}~\cite{Geant4} simulation to propagate the particles emitted by the decays and record the energy released in the detector. 
The \textit{Arby} output is then processed to incorporate the detector features, such as time and energy resolution, getting in the end a simulated spectrum for any radioisotope. 

To validate the simulation models, we performed benchmark measurements with calibration sources containing isotopes of certified activity and emitting $\gamma$-rays in a wide energy range from 59~keV to 1408~keV ($^{241}$Am, $^{137}$Cs, $^{60}$Co, $^{133}$Ba, $^{152}$Eu).
The activities reconstructed from the analysis and simulation of experimental measurements were compared to the certified ones, obtaining an agreement within 5\% for all configurations with the sources put at different distances from the detector. 

After irradiations, we performed $\gamma$-spectroscopy measurements of activated samples for about one month, tuning the measurement and waiting times in order to observe the $\gamma$-lines of both short-lived and long-lived isotopes. Most samples were measured on both HPGe detectors, to achieve a better control of systematic uncertainty in activation rate evaluation.

\section{Data analysis}

In this section we present at first the analysis of Al-Co activation monitors, then the characterization of the neutron flux spatial distribution, and finally the activation rate measurement of the different reactions which will then be used for neutron flux spectrum unfolding.

\begin{figure*}[htb!]
\centering
\subfloat{\includegraphics[width=0.9\textwidth]{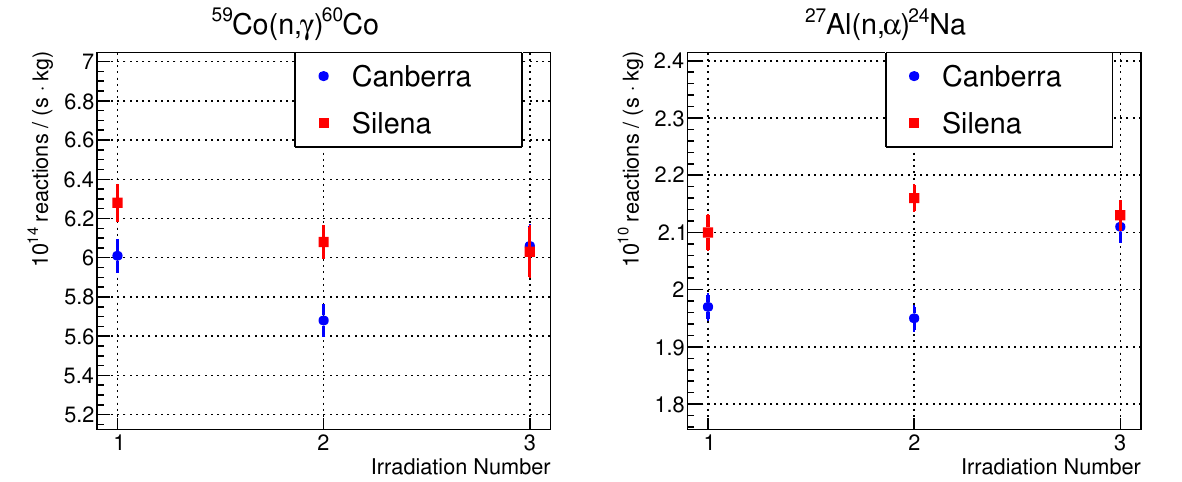}}\\
\subfloat{\includegraphics[width=0.9\textwidth]{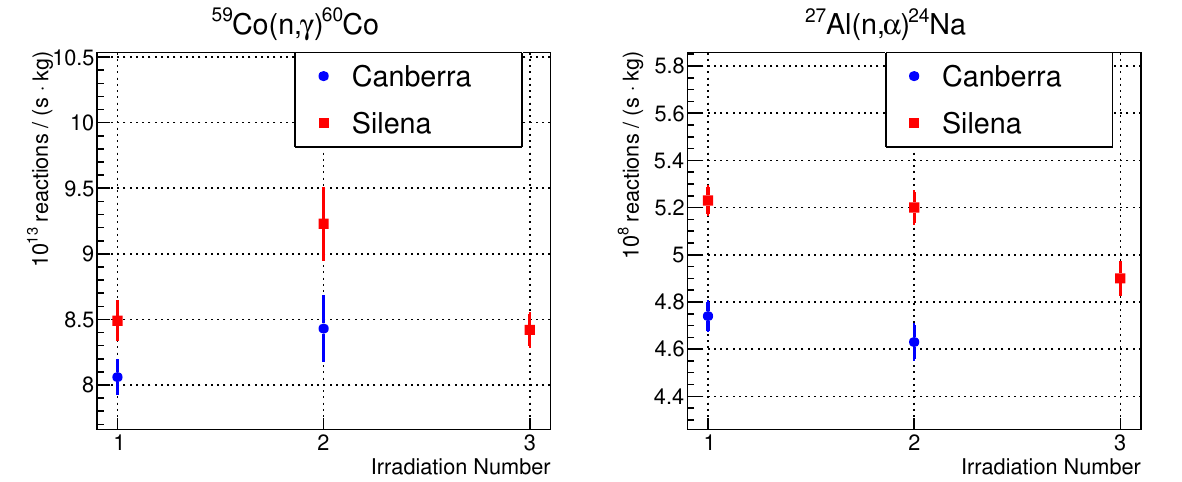}}
\caption{Results from Al-Co activation monitors positioned in Central Channel (top) and in Lazy Susan -- channel \#1 (bottom) facilities to check the repeatability of the experimental conditions throughout the different irradiations. The plots at left show the measured activation rate of $^{59}$Co$(n,\gamma)^{60}$Co reaction, whereas at right the corresponding results for $^{27}$Al$(n,\alpha)^{24}$Na reaction are shown. The blue and red colors are used to differentiate the results obtained from the experimental measurements performed with the Canberra and Silena HPGe detectors, respectively.}
\label{Fig:AlCoMonitor}
\end{figure*}

\subsection{Analysis of activation monitors}

To assess the repeatability of the activation measurements in the three different irradiations, we placed Al-Co monitor samples in the same positions at the bottom of Central Channel and Lazy Susan (channel \#1) holders.
In stable conditions, we expect that the activation rates of $^{59}$Co$(n,\gamma)^{60}$Co and $^{27}$Al$(n,\alpha)^{24}$Na reactions are compatible among different irradiations.
Particularly, we use the $(n,\gamma)$ reaction on cobalt to monitor the neutron flux in the thermal and intermediate range, and the $(n,\alpha)$ reaction on aluminum to measure the flux of fast neutrons with energy $>5$~MeV.

\begin{figure*}[!htb]
\centering
\subfloat{\includegraphics[width=0.9\textwidth]{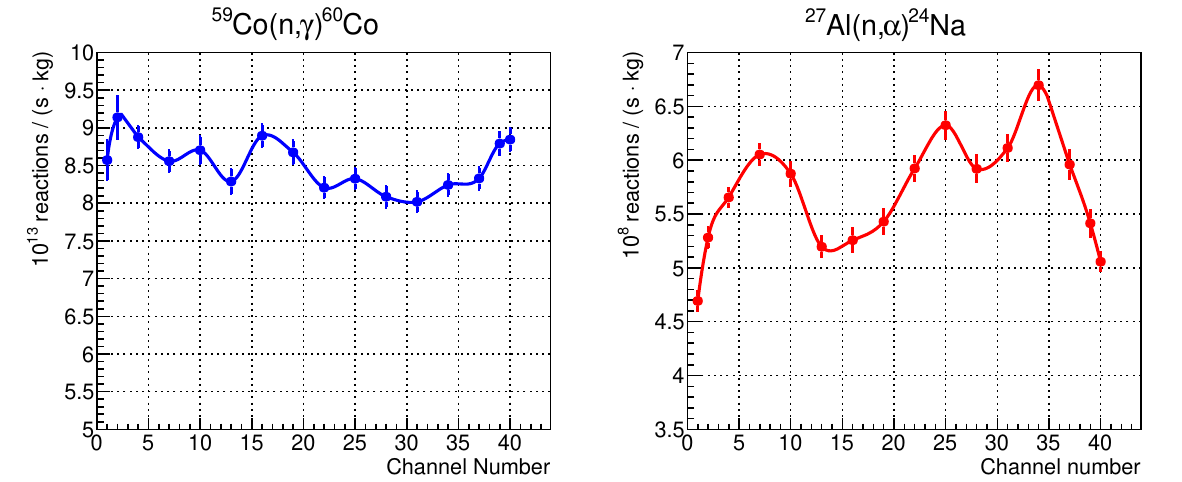}}\\
\subfloat{\includegraphics[width=0.9\textwidth]{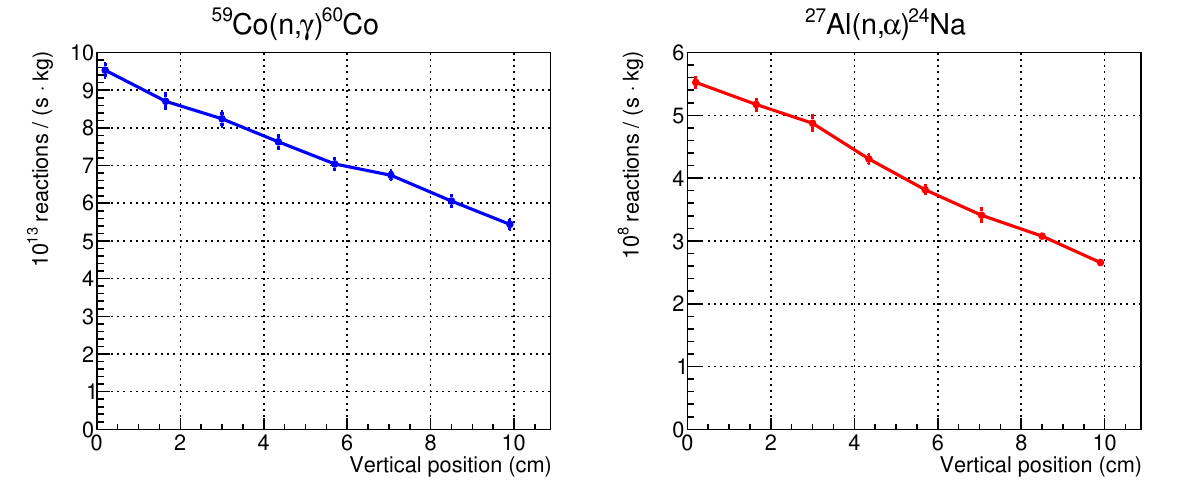}}\\
\caption{Top: results of Al-Co activation samples positioned in the bottom of Lazy Susan channels (about 20~cm above the core mid-plane) to map the flux in the annulus around the core. Bottom: results of Al-Co activation samples positioned at different heights in the channel \#40 of Lazy Susan facility to map the vertical gradient of the flux (the zero of the x-axis is at the bottom of the irradiation holder, i.e. the lowest sample position in the Lazy Susan facility).}
\label{Fig:MapLS}
\end{figure*}

All monitor samples, except one, were measured on both HPGe detectors (Silena and Canberra). 
By separately analyzing the Silena and Canberra data series shown in Fig.~\ref{Fig:AlCoMonitor}, we observe that the activation rates are compatible within a $\pm5$\% range, proving that the neutron flux does not significantly vary from an irradiation to another. 
In Fig.~\ref{Fig:AlCoMonitor} it can also be noted that the activation rates obtained from Silena detector measurements are on average higher than those referring to the same samples measured on Canberra one. This is due to the fact that the detector efficiency estimated by MC simulations is affected by systematic uncertainty up to $\sim5$\%. 
Conceding that the activation rate measurements with the same detector are affected by a bias of this entity, the Silena and Canberra data series return to be compatible. 
This point is important for the rest of the analysis, because we have to propagate this systematic uncertainty when using the activation rate data to unfold the neutron flux. 
Obviously, this error is reduced when averaging the results from two independent measurements on different HPGe detectors.

\subsection{Characterization of neutron flux spatial distribution}

\begin{figure*}[!htb]
\centering
\includegraphics[width=0.9\textwidth]{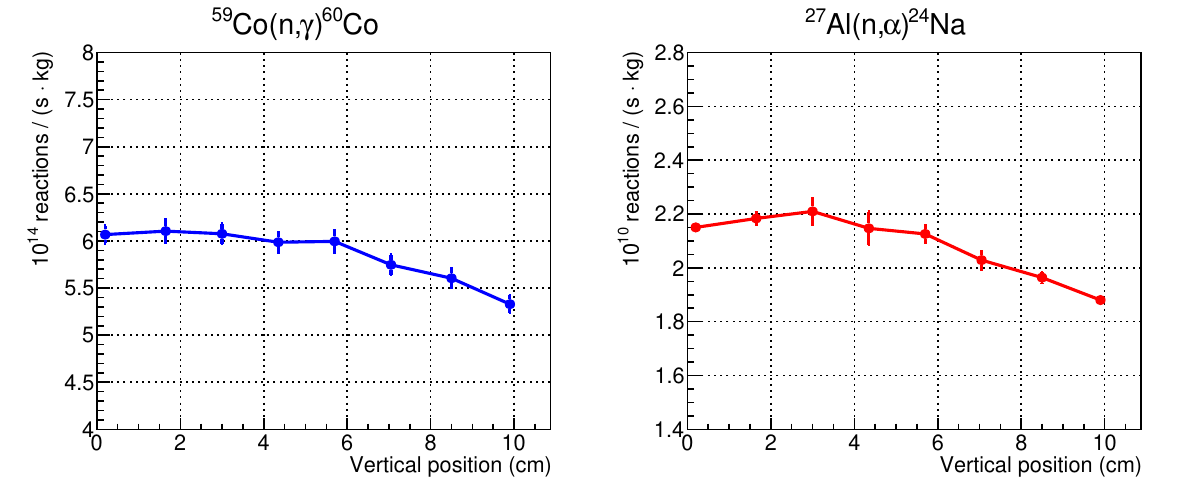}
\caption{Results of Al-Co activation samples positioned at different heights in the Central Channel to map the vertical gradient of the flux (the zero of the x-axis corresponds to the bottom of the Central Channel irradiation holder, positioned at core mid-plane).}
\label{Fig:MapCC}
\end{figure*}

To map the spatial profile of the neutron flux intensity inside the reactor facilities, we activated several Al-Co samples in the annulus formed by the 40 channels of Lazy Susan specimen rack, and along the vertical axes of Central Channel and Lazy Susan irradiation holders. 
This technique, as mentioned above, allows to simultaneously probe the neutron flux in both the thermal-intermediate and in the fast regions. 
Therefore, the activation rate profiles shown in Figs.~\ref{Fig:MapLS} and \ref{Fig:MapCC} correspond, except for a constant of proportionality, to the neutron flux spatial distributions.

The non-uniformity of the flux in the different positions of the Lazy Susan circular rack is a known aspect, related to the asymmetric core loading. The experimental results of the thermal-intermediate neutron flux component exhibit maximum differences of $\pm$7\% with respect to the central value, whereas the fast flux distribution is affected by higher variations, up to $\pm$18\% (Fig.~\ref{Fig:MapLS} (top)). It is worth noting that the lowest fast flux values are observed in the channels adjacent to the graphite \textit{dummy} elements (see the core scheme in Fig.~\ref{Fig:TRIGA}).

In Fig.~\ref{Fig:MapLS} (bottom), we show the vertical distribution of the neutron flux in the channel \#40 of Lazy Susan. Since this facility is positioned in the upper part of the reflector, about 20~cm above mid-plane, the flux shows a decreasing gradient in both thermal-intermediate and fast components.

In Fig.~\ref{Fig:MapCC}, we show the vertical distribution of the neutron flux in the Central Channel. The flux is almost uniform in the lower 6~cm, and slightly decreases at higher distances from the bottom position of the holder.

The experimental results highlight that the neutron flux is not spatially homogeneous, especially in the Lazy Susan facility. In this case, we also observe that the ratio of thermal-intermediate versus fast flux component is not constant. This implies that the energy spectrum varies as function of both vertical and longitudinal coordinates in the Lazy Susan rack.
If high accuracy is required in the evaluation of fluence as function of neutron energy in all channels of Lazy Susan facility, a further characterization analysis is needed, because the data collected in this experimental campaign allow for the spectrum unfolding in one Lazy Susan channel (namely channel \#1, where the samples containing different elements have been irradiated). Monte Carlo (MC) simulations are a suitable tool for this purpose. Indeed, a MC model for neutronics simulations of the TRIGA RC-1 reactor could be developed and validated using the experimental data reported here as a benchmark, and subsequently exploited to estimate the neutron flux spectrum in all irradiation positions.

\subsection{Activation rate results}

\renewcommand{\arraystretch}{1.1}
\begin{table*}[htb!]
\begin{center}
\begin{tabular}{l|c|c}
\multirow{2}{*}{Reaction}  & Lazy Susan (ch. \#1) & Central Channel \\
                           & SSA (Bq/g) & SSA (Bq/g) \\
\hline
\rule{0pt}{3ex}
$^{45}$Sc $(n,\gamma)$ $^{46}$Sc	&	$(8.65\pm0.49)\times10^{10}$	&	$(5.64\pm0.26)\times10^{11}$	\\
$^{50}$Cr $(n,\gamma)$ $^{51}$Cr 	&	$(1.58\pm0.06)\times10^{9}$	&	$(1.13\pm0.04)\times10^{10}$	\\
$^{58}$Fe $(n,\gamma)$ $^{59}$Fe	&	$(8.34\pm0.46)\times10^{6}$	&	$(5.67\pm0.31)\times10^{7}$	\\
$^{59}$Co $(n,\gamma)$ $^{60}$Co	&	$(8.47\pm0.30)\times10^{10}$	&	$(6.51\pm0.22)\times10^{11}$	\\
$^{64}$Ni $(n,\gamma)$ $^{65}$Ni	&	$(3.15\pm0.11)\times10^{7}$	&	$(2.27\pm0.08)\times10^{8}$	\\
$^{64}$Zn $(n,\gamma)$ $^{65}$Zn	&	$(3.42\pm0.14)\times10^{8}$	&		\\
$^{71}$Ga $(n,\gamma)$ $^{72}$Ga 	&	$(4.18\pm0.15)\times10^{9}$	&	$(3.82\pm0.13)\times10^{10}$	\\
$^{74}$Ge $(n,\gamma)$ $^{75}$Ge	&	$(3.05\pm0.19)\times10^{8}$	&		\\
$^{76}$Ge $(n,\gamma)$ $^{77}$As	&	$(2.17\pm0.11)\times10^{7}$	&	$(2.07\pm0.10)\times10^{8}$	\\
$^{75}$As $(n,\gamma)$ $^{76}$As	&	$(9.78\pm0.35)\times10^{9}$	&	$(1.09\pm0.04)\times10^{11}$	\\
$^{74}$Se $(n,\gamma)$ $^{75}$Se 	&	$(9.72\pm0.48)\times10^{8}$	&	$(1.01\pm0.05)\times10^{10}$	\\
$^{82}$Se $(n,\gamma)$ $^{83}$Se	&	$(5.28\pm0.23)\times10^{6}$	&		\\
$^{85}$Rb $(n,\gamma)$ $^{86}$Rb 	&	$(7.93\pm0.31)\times10^{8}$	&	$(9.66\pm0.36)\times10^{9}$	\\
$^{94}$Zr $(n,\gamma)$ $^{95}$Zr	&	$(1.46\pm0.06)\times10^{7}$	&	$(1.29\pm0.04)\times10^{8}$	\\
$^{96}$Zr $(n,\gamma)$ $^{97}$Zr 	&	$(7.20\pm0.67)\times10^{6}$	&	$(1.24\pm0.04)\times10^{8}$	\\
$^{98}$Mo $(n,\gamma)$ $^{99}$Mo	&	$(1.09\pm0.04)\times10^{8}$	&	$(1.63\pm0.06)\times10^{9}$	\\
$^{109}$Ag $(n,\gamma)$ $^{110m}$Ag	&	$(7.13\pm0.51)\times10^{10}$	&	$(7.74\pm0.55)\times10^{11}$	\\
$^{114}$Cd $(n,\gamma)$ $^{115}$Cd	&	$(2.51\pm0.12)\times10^{8}$	&	$(3.51\pm0.17)\times10^{9}$	\\
$^{113}$In $(n,\gamma)$ $^{114m}$In	&	$(1.31\pm0.10)\times10^{9}$	&	$(2.08\pm0.13)\times10^{9}$	\\
$^{115}$In $(n,\gamma)$ $^{116m}$In 	&	$(3.45\pm0.18)\times10^{11}$	&		\\
$^{121}$Sb $(n,\gamma)$ $^{122}$Sb	&	$(7.39\pm0.26)\times10^{9}$	&	$(1.03\pm0.03)\times10^{11}$	\\
$^{123}$Sb $(n,\gamma)$ $^{124}$Sb	&	$(3.36\pm0.12)\times10^{9}$	&	$(4.60\pm0.15)\times10^{10}$	\\
$^{139}$La $(n,\gamma)$ $^{140}$La	&	$(8.62\pm0.71)\times10^{9}$	&	$(6.59\pm0.22)\times10^{10}$	\\
$^{151}$Eu $(n,\gamma)$ $^{152}$Eu 	&	$(4.04\pm0.14)\times10^{12}$	&	$(2.71\pm0.09)\times10^{13}$	\\
$^{153}$Eu $(n,\gamma)$ $^{154}$Eu	&	$(1.70\pm0.06)\times10^{11}$	&	$(1.36\pm0.05)\times10^{12}$	\\
$^{159}$Tb $(n,\gamma)$ $^{160}$Tb	&	$(3.20\pm0.19)\times10^{10}$	&	$(3.37\pm0.15)\times10^{11}$	\\
$^{176}$Lu $(n,\gamma)$ $^{177}$Lu	&	$(7.91\pm0.77)\times10^{10}$	&	$(6.64\pm0.41)\times10^{11}$	\\
$^{186}$W $(n,\gamma)$ $^{187}$W	&	$(1.19\pm0.04)\times10^{10}$	&	$(1.20\pm0.04)\times10^{11}$	\\
$^{191}$Ir $(n,\gamma)$ $^{192}$Ir	&	$(2.52\pm0.09)\times10^{11}$	&	$(2.01\pm0.07)\times10^{12}$	\\
$^{193}$Ir $(n,\gamma)$ $^{194}$Ir	&	$(6.57\pm0.24)\times10^{10}$	&	$(6.80\pm0.23)\times10^{11}$	\\
$^{197}$Au $(n,\gamma)$ $^{198}$Au	&	$(8.81\pm0.30)\times10^{10}$	&	$(9.30\pm0.30)\times10^{11}$	\\
$^{232}$Th $(n,\gamma)$ $^{233}$Pa	&	$(5.44\pm0.20)\times10^{9}$	&	$(5.70\pm0.19)\times10^{10}$	\\
$^{238}$U $(n,\gamma)$ $^{239}$Np	&	$(6.35\pm0.27)\times10^{9}$	&	$(8.63\pm0.35)\times10^{10}$	\\
\hline	
\rule{0pt}{3ex}	
$^{24}$Mg $(n,p)$ $^{24}$Na	&	$(9.24\pm0.36)\times10^{5}$	&	$(3.63\pm0.12)\times10^{7}$	\\
$^{27}$Al $(n,\alpha)$ $^{24}$Na	&	$(4.98\pm0.17)\times10^{5}$	&	$(2.04\pm0.07)\times10^{7}$	\\
$^{54}$Fe $(n,p)$ $^{54}$Mn	&	$(1.50\pm0.07)\times10^{6}$	&	$(6.00\pm0.23)\times10^{7}$	\\
$^{56}$Fe $(n,p)$ $^{56}$Mn	&	$(3.86\pm0.14)\times10^{5}$	&	$(1.32\pm0.05)\times10^{7}$	\\
$^{58}$Ni $(n,p)$ $^{58}$Co	&	$(2.15\pm0.07)\times10^{7}$	&	$(8.95\pm0.29)\times10^{8}$	\\
$^{58}$Ni $(n,pn)$ $^{57}$Co	&	$(6.07\pm0.31)\times10^{4}$	&	$(2.15\pm0.07)\times10^{6}$	\\
$^{60}$Ni $(n,p)$ $^{60}$Co	&	$(2.08\pm0.16)\times10^{5}$	&	$(7.51\pm0.25)\times10^{6}$	\\
$^{115}$In $(n,n')$ $^{115m}$In  	&	$(2.77\pm0.09)\times10^{7}$	&	$(1.28\pm0.04)\times10^{9}$	\\
\end{tabular}
\end{center}
\caption{Specific saturation activities (SSA) of radiative capture (top part) and threshold (bottom part) neutron-induced reactions measured in the Lazy Susan (channel \#1) and in the Central Channel of TRIGA RC-1 reactor.}
\label{Tab:ActRate}
\end{table*}

To measure the activation rates of a large set of neutron-induced reactions (later used for flux spectrum unfolding), we organized the data analysis according to the following steps.
\begin{enumerate}
\item We analyze the lines in the $\gamma$-spectroscopy measurements to identify the activated isotopes.
\item For each activated isotope and measurement configuration (i.e. HPGe detector and sample position), we run a MC simulation to get the detection efficiency of each $\gamma$-line.
\item When an isotope produces more than one $\gamma$-line, we check that the relative intensity of the measured peaks is compatible with that predicted by simulations. This check allows to easily identify overlays of $\gamma$-lines produced by other activated isotopes in the same sample or natural background.
\item We calculate the isotope activity from the number of counts in each $\gamma$-line. In case of isotopes producing two or more $\gamma$-lines, we compute the weighted average to get the best estimate of the activity.
\item Finally, we use Eq.~\ref{Eq:Activity} to get the activation rate $R$, and we average the results from the measurements on the two HPGe detectors (Silena and Canberra) to get the best estimate of $R$ for each observed reaction.
\end{enumerate}
As shown before in Fig.~\ref{Fig:MapLS}, the activation rates in the Lazy Susan facility exhibit a not negligible gradient along the vertical direction. Since the samples were stacked in the irradiation holder at different heights (up to about 6~cm from the holder bottom), we bring back the results to a common reference position by applying a correction factor derived from the Al-Co samples analysis. Particularly, we use the $^{59}$Co$(n,\gamma)^{60}$Co activation rate gradient to correct radiative capture activation rates, and $^{27}$Al$(n,\alpha)^{24}$Na one for threshold reactions. Such correction is not needed for Central Channel data, because the characterization performed through Al-Co monitors points out a negligible gradient within the first 6~cm from the bottom.

Overall, we measured 33 different $(n,\gamma)$ reactions and 8 threshold reactions.
The activation rates divided by the mass of the target element in the sample are reported in Tab.~\ref{Tab:ActRate}. These results correspond to the \textit{specific saturation activities} (SSA), a quantity most commonly used in neutron activation analysis.
To list the reactions, we use a notation in which the target nucleus appears on the left, the reaction in the center and the measured radioisotope on the right. For example, the $(n,\gamma)$ reaction on $^{238}$U is measured by analyzing the $\gamma$-lines emitted by $^{239}$Np, which is produced after the decay of the relatively short-lived $^{239}$U.
When a reaction proceeds through different branches, activating both the ground and the metastable state of an isotope, and only one can be measured, we quote the total reaction rate, obtained after dividing by the branching ratio of the observed reaction channel.

The uncertainties associated to SSA results listed in Tab.~\ref{Tab:ActRate} include: the statistical (Poisson) uncertainty on $\gamma$-line counts in measured and simulated spectra, the vertical gradient correction factor uncertainty (Lazy Susan data only), the branching ratio uncertainty, the target nucleus isotopic abundance uncertainty, and the systematic uncertainty on absolute activity reconstruction by $\gamma$-spectroscopy analysis.
Uncertainties on element mass and on the duration of irradiations and measurements are negligible.

To assess the systematic uncertainty, we analyzed the results of $^{60}$Co activity reconstruction in 8 Al-Co samples that have been measured in practically identical conditions on both HPGe detectors. For each couple of measurements ($A_i$), we calculated the two residuals ($\delta_i$) from the weighted mean activity value ($\overline{A}$):

\begin{equation}
\delta_i=\dfrac{A_i-\overline{A}}{\overline{A}} \quad i = \text{Canberra}, \text{Silena}
\end{equation}
Then, we computed the mean value ($\mu_i$) of each $\delta_i$ data set, obtaining (-2.0$\pm$0.3)\% and (+2.6$\pm$0.5)\% for Canberra and Silena detectors, respectively. The difference of $\mu_i$ provides an estimate of the systematic uncertainty affecting the absolute activity reconstruction when a single HPGe detector is used. 
When a sample is measured on two detectors, the systematic uncertainty on the combined activity evaluation is reduced by a factor $\sqrt{2}$, thus in our case we propagate a $\pm$3.2\% systematic error.

\section{Neutron flux unfolding and results}
The neutron flux spectrum unfolding is performed by solving a system of linear equations derived from Eq.~\ref{Eq:ActRate} after discretizing the energy spectrum into $n$ groups:

\begin{equation}
\label{Eq:ActRateDiscrete}
\dfrac{R_j}{\mathcal{N}_j} =  \sum_{i=0}^{n} \sigma_{ij} \phi_i \quad \left(\phi_i \equiv \int_{E_i}^{E_{i+1}} \phi(E)\,dE \right) 
\end{equation}
where $j$ is an index to label the different reactions, $\phi_i$ is the neutron flux intensity in the $i$-th group, and $\sigma_{ij}$ is the \textit{group effective cross section}, which is defined as:

\begin{equation}
\label{Eq:XSeff}
\sigma_{ij} = \dfrac{\int_{E_i}^{E_{i+1}} \sigma_j(E)\,\phi(E)\,dE}{\int_{E_i}^{E_{i+1}} \phi(E)\,dE}
\end{equation}
and corresponds to the average cross section, weighted on neutron spectrum, in the $i$-th energy group.

The system in Eq.~\ref{Eq:ActRateDiscrete} is solved with a Bayesian statistical approach that allows to propagate the experimental uncertainties affecting the parameters ($R_j / \mathcal{N}_j$ and $\sigma_{ij}$), and to select the physical solutions for $\phi_i$ unknown variables, which are positive definite.
Practically, we sample the \textit{joint posterior} PDF (Probability Density Function) $p(\phi_{i}|R_j,\sigma_{ij})$ exploiting the JAGS tool~\cite{JAGS,JAGS_manual}, which uses the Markov Chains Monte Carlo (MCMC) method~\cite{Gelman}. 
At the end, by marginalizing the joint PDF, we get the posterior PDF for each flux group $\phi_i$ and the correlations among them. The averages and standard deviations of $\phi_i$ marginalized PDFs are calculated to provide a multi-group neutron flux measurement, which represents the result of the spectrum unfolding.

More details about the Bayesian analysis that we use to unfold the neutron spectrum can be found in Refs.~\cite{BayesianSpectrum,ChipIR}. 
We recall here two relevant aspects regarding this unfolding method. 

The first one is that the choice of the binning used to discretize the energy spectrum into $n$ groups must take into account the characteristics of the activation cross sections, and in particular their energy dependence. 
To avoid getting an indeterminate system of equations, the number of groups and their ranges must be chosen in such a way as to ensure that, for each group, there is at least one reaction induced in a non-negligible percentage by neutrons in that energy range. 
Moreover, we cannot split into more groups the energy ranges where neutron cross sections exhibit the same energy dependence. This is the case of the thermal range, in which the radiative capture cross section is proportional to $1/\sqrt{E}$ for all reactions. 
According to these criteria, in the intermediate region we define energy bins that include at least one of the main resonances of neutron capture cross sections, whereas in the fast region we set bin edges in correspondence with the energy thresholds of some measured reactions. For this analysis we use $n=10$ energy groups.

The second aspect to be recalled is that the unfolding procedure requires a \textit{guess spectrum} to calculate effective cross sections $\sigma_{ij}$ in each energy group. In this regard, we point out that even if the unfolding results have some degree of dependence on the \textit{intra-group} spectrum shape used for $\sigma_{ij}$ calculation, there is no constraint on $\phi_{i}$ intensities. 

\begin{figure}[htb!]
\centering
\includegraphics[width=\linewidth]{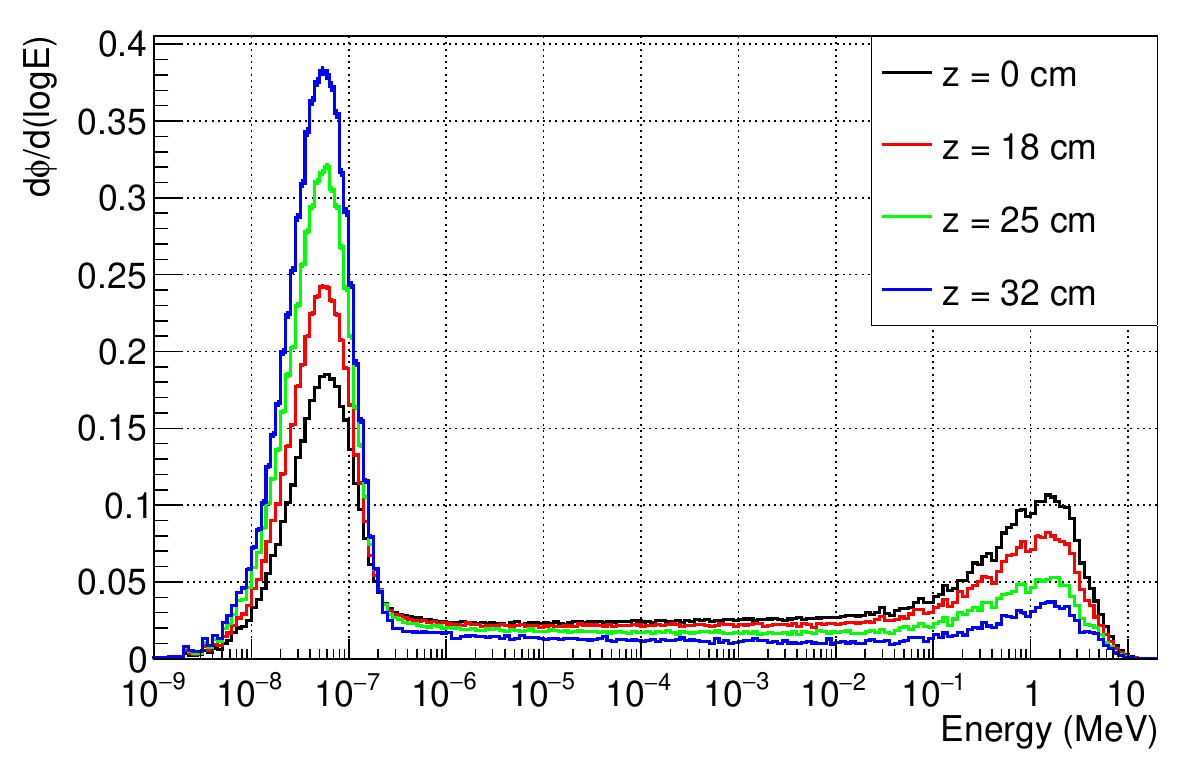}
\caption{Some of the spectra (shown in lethargic scale and normalized to unity) belonging to the set of guess spectra used for the unfolding procedure. In particular, these spectra have been generated using the MCNP model of the TRIGA Mark II reactor in Pavia, simulating the neutron flux in the Rabbit Channel at different heights above the core mid-plane.}
\label{Fig:GuessSpectra}
\end{figure}

\renewcommand{\arraystretch}{1.2}
\begin{table*} [htb!]
\begin{minipage}[b]{0.5\linewidth}
\begin{center}
\begin{tabular}{c|c|c}
Energy range	&	Neutron flux	&	Relative	\\
(min -- max)	    &	[n/(cm$^2$s)]	&	uncertainty	\\
\hline					
\rule{0pt}{3ex}
(0.001 -- 0.251) eV	&	$(2.53\pm0.04)\times10^{11}$	&	1.6\%	\\
(0.251 -- 1.995) eV	&	$(1.78\pm0.24)\times10^{10}$	&	13\%	\\
(1.995 -- 10) eV 	&	$(8.55\pm0.85)\times10^{9}$	&	9.9\%	\\
(10 -- 35.5) eV  	&	$(1.10\pm0.09)\times10^{10}$	&	8.5\%	\\
(35.5 -- 100) eV	&	$(5.90\pm1.40)\times10^{9}$	&	24\%	\\
(100 -- 316) eV   &	$(8.63\pm0.84)\times10^{9}$	&	9.7\%	\\
(316 -- 5$\times10^{5}$) eV   &	$(5.36\pm0.59)\times10^{10}$	&	11\%	\\
(0.5 -- 5) MeV  	&	$(3.38\pm0.21)\times10^{10}$	&	6.2\%	\\
(5 -- 10) MeV	    &	$(1.50\pm0.07)\times10^{9}$	&	4.3\%	\\
(10 -- 20) MeV	&	$(5.48\pm0.55)\times10^{7}$	&	10\%	\\
\hline	
\rule{0pt}{3ex}				
Total	&	$(3.93\pm0.06)\times10^{11}$	&	1.6\%	\\
\end{tabular}
\end{center}
\caption{Results of multi-group neutron flux measurement in the Lazy Susan (channel \#1) facility at 100~kW power.}
\label{Tab:LSunfolded}
\end{minipage}\hfill
\begin{minipage}[b]{0.42\linewidth}
\centering
\includegraphics[width=\linewidth]{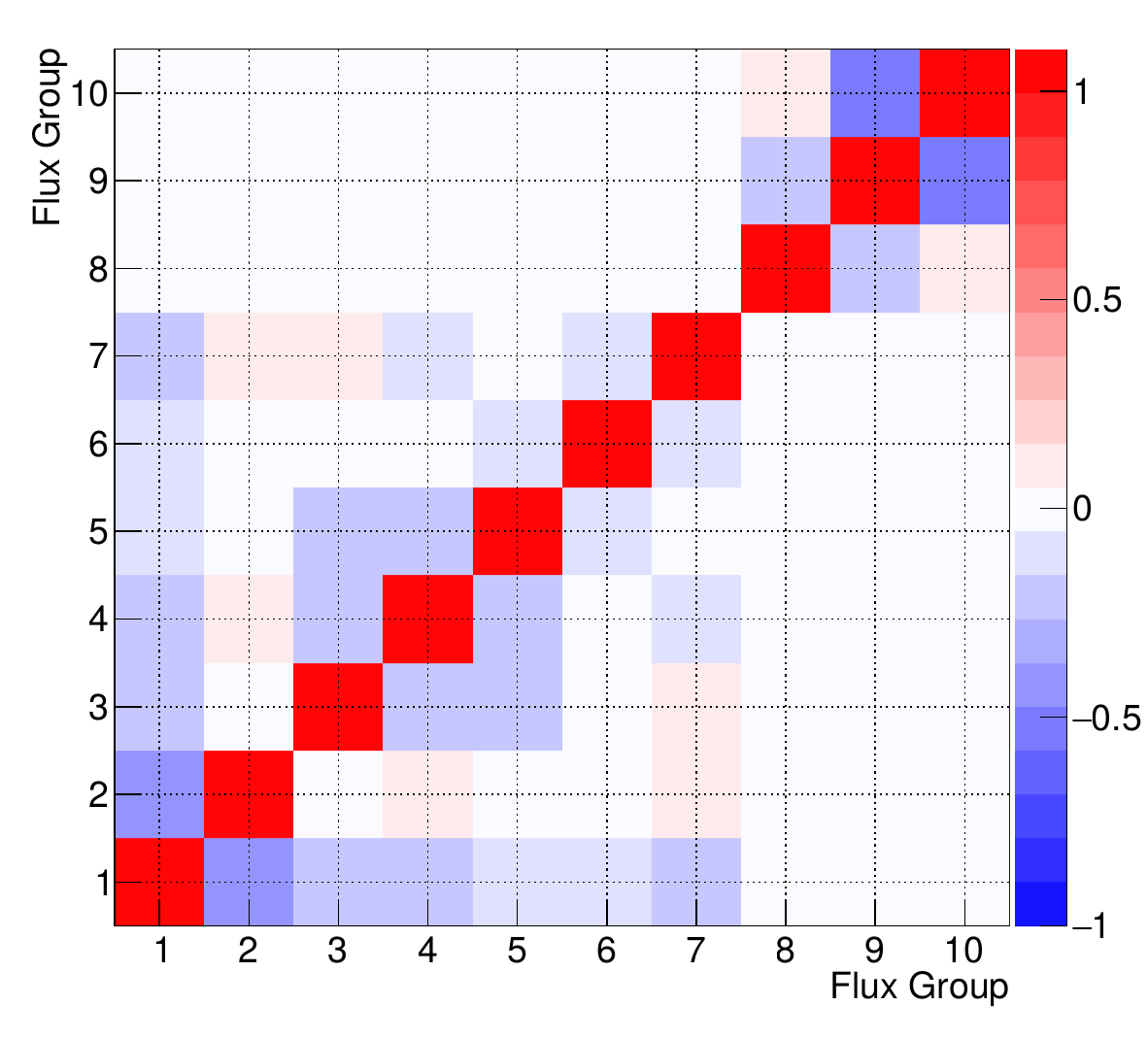}
\captionof{figure}{Correlations among groups used to unfold the neutron flux spectrum in Lazy Susan facility. }
\label{Fig:LSmatrix}
\end{minipage}
\vskip\baselineskip
\begin{minipage}[b]{\linewidth}
\centering
\includegraphics[width=0.75\textwidth]{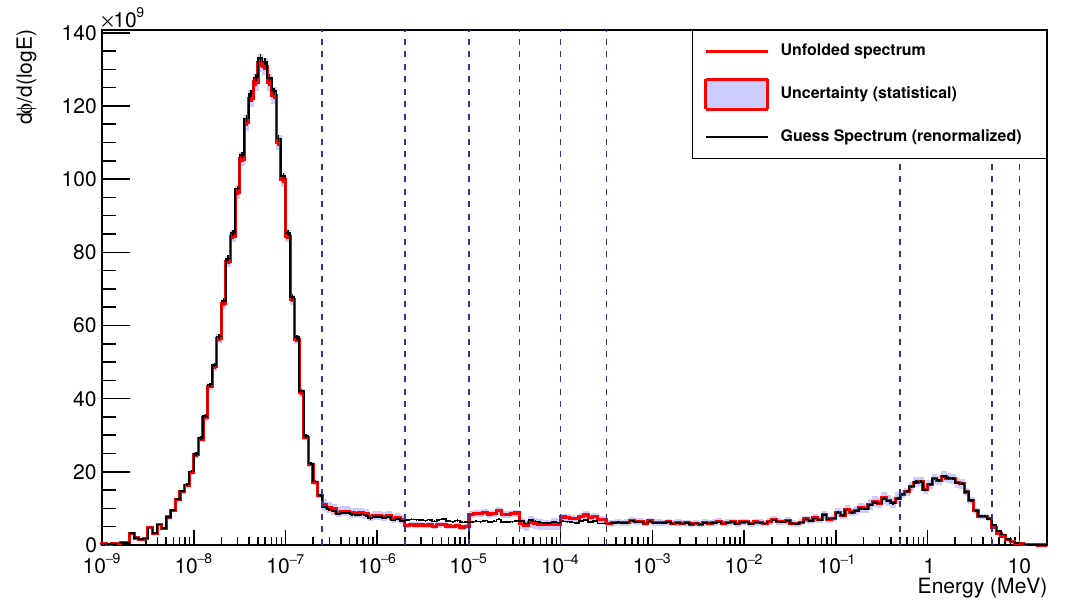}
\captionof{figure}{Unfolded neutron flux spectrum in the Lazy Susan (channel \#1) facility at 100~kW power. The flux spectrum is represented using the lethargic scale, with flux in units of n/(cm$^{2}$s). The red line is obtained by normalizing the guess spectrum groups to the multi-group flux intensities reported in Tab.~\ref{Tab:LSunfolded}. The subdivision of energy ranges is shown by the dotted vertical lines. The light blue shaded area corresponds to the uncertainty of the flux in each group. The black line is the guess spectrum used for the unfolding normalized to the total flux intensity.}
\label{Fig:LSunfolded}
\end{minipage}
\end{table*}

\renewcommand{\arraystretch}{1.2}
\begin{table*} [htb!]
\begin{minipage}[b]{0.5\linewidth}
\begin{center}
\begin{tabular}{c|c|c}
Energy range	&	Neutron flux	&	Relative	\\
(min -- max)	    &	[n/(cm$^2$s)]	&	uncertainty	\\
\hline					
\rule{0pt}{3ex}
(0.001 -- 0.251) eV	&	$(1.85\pm0.03)\times10^{12}$	&	1.8\%	\\
(0.251 -- 1.995) eV	&	$(2.33\pm0.35)\times10^{11}$	&	15\%	\\
(1.995 -- 10) eV 	&	$(1.59\pm0.09)\times10^{11}$	&	5.8\%	\\
(10 -- 35.5) eV  	&	$(1.76\pm0.10)\times10^{11}$	&	5.6\%	\\
(35.5 -- 100) eV	&	$(1.18\pm0.15)\times10^{11}$	&	13\%	\\
(100 -- 316) eV	&	$(1.43\pm0.06)\times10^{11}$	&	4.3\%	\\
(316 -- 5$\times10^{5}$) eV	&	$(1.29\pm0.09)\times10^{12}$	&	6.9\%	\\
(0.5 -- 5) MeV  	&	$(1.29\pm0.07)\times10^{12}$	&	5.6\%	\\
(5 -- 10) MeV	&	$(5.92\pm0.25)\times10^{10}$	&	4.2\%	\\
(10 -- 20) MeV	&	$(1.90\pm0.19)\times10^{9}$	&	10\%	\\
\hline	
\rule{0pt}{3ex}				
Total	&	$(5.32\pm0.11)\times10^{12}$	&	2.1\%	\\
\end{tabular}
\end{center}
\caption{Results of multi-group neutron flux measurement in Central Channel at 100~kW power.}
\label{Tab:CCunfolded}
\end{minipage}\hfill
\begin{minipage}[b]{0.42\linewidth}
\centering
\includegraphics[width=\linewidth]{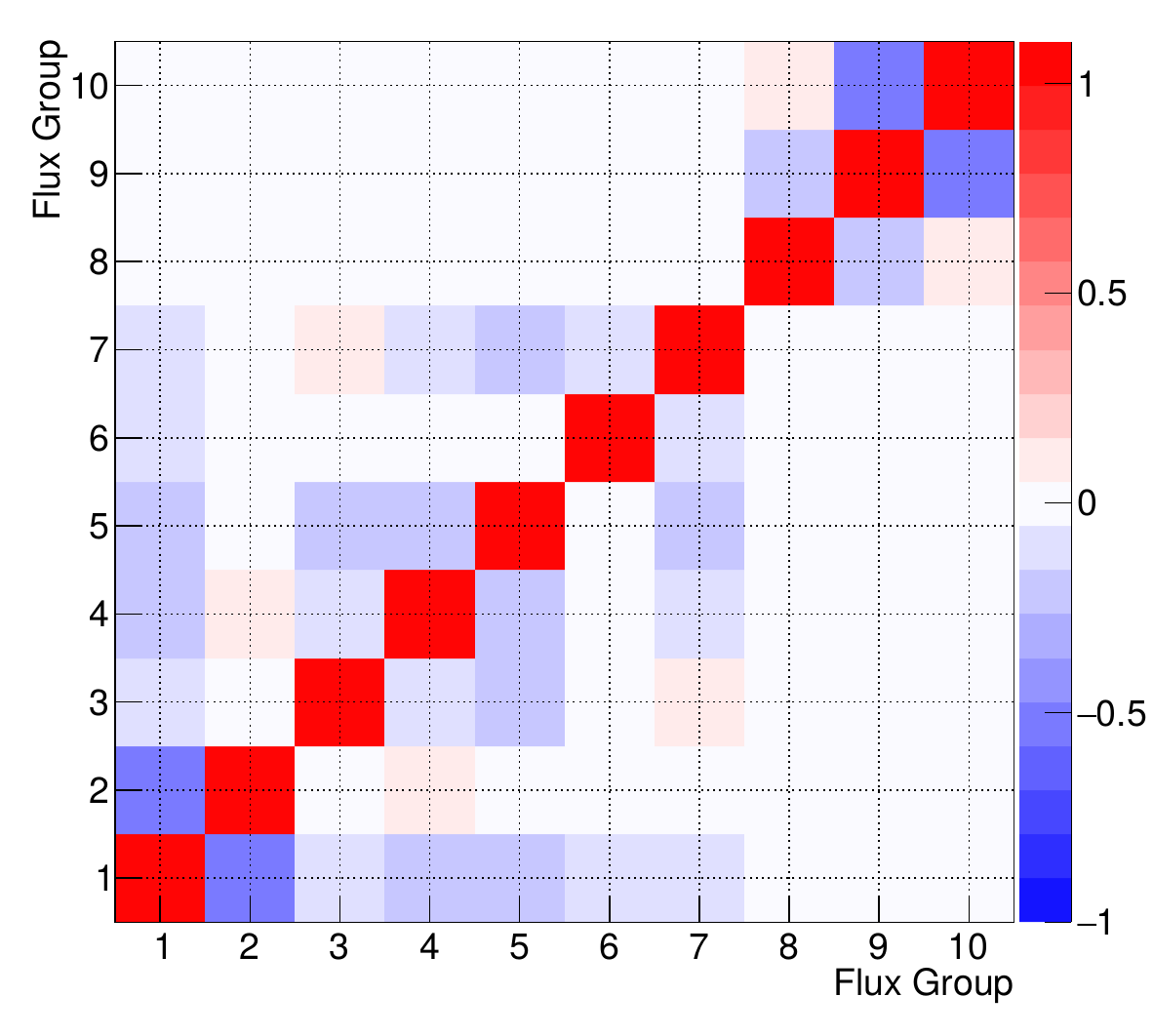}
\captionof{figure}{Correlations among groups used to unfold the neutron flux spectrum in Central Channel.}
\label{Fig:CCmatrix}
\end{minipage}
\vskip\baselineskip
\begin{minipage}[b]{\linewidth}
\centering
\includegraphics[width=0.75\textwidth]{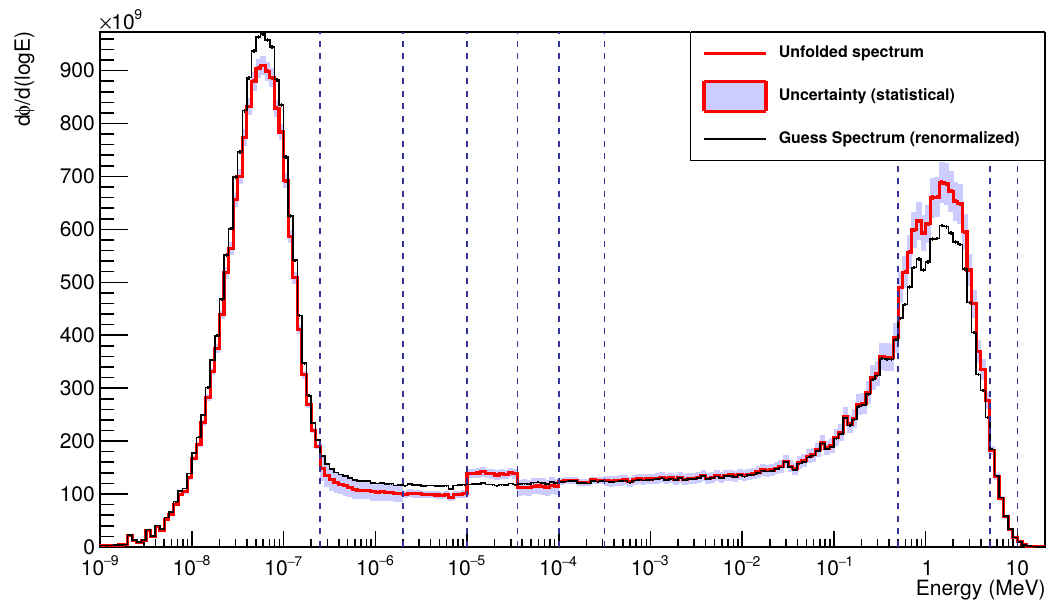}
\captionof{figure}{Unfolded neutron flux spectrum in Central Channel at 100~kW power (same plot description as Fig.~\ref{Fig:LSunfolded}.)}
\label{Fig:CCunfolded}
\end{minipage}
\end{table*}

To select a suitable guess spectrum, we exploit a set of neutron spectra generated by means of a Monte Carlo simulation model (based on the MCNP code~\cite{mcnp}), developed a few years ago for the neutronics analysis of the TRIGA Mark~II reactor in Pavia (Italy)~\cite{FirstReactorMODEL,FreddoPulito,CaldoPulito,Burnup}. This set includes about a hundred spectra obtained by tallying the neutron flux in various positions of the reactor core, reflector and irradiation facilities. 
In thermal reactors, like TRIGA ones, the neutron spectrum has three main components characterized by well-defined energy dependences~\cite{lamarsh}. Indeed, fast neutrons from fissions are produced with an energy spectrum usually referred to as Watt distribution~\cite{watt}, and are subsequently moderated to thermal energies, where they exhibit a Maxwell distribution. The moderation process generates a neutron spectrum proportional to $E^{-\beta}$ (with $\beta \simeq 1$) in the intermediate range.
As shown in Fig.~\ref{Fig:GuessSpectra}, these features are common to all neutron spectra sampled at different reactor positions. What changes is the ratio between thermal and fast flux components. Thus, as a first approximation, we can calculate $\sigma_{ij}$ values by using as guess any of the spectra belonging to the aforementioned set.

To calculate the effective cross sections we use the data published in the ENDF/B-VII.1 libraries~\cite{ENDF}, and we take the uncertainties of thermal neutron capture cross sections and resonance integrals from the BNL-98403-2012-JA Report~\cite{PRITYCHENKO20123120}, based on the Low Fidelity Covariance Project~\cite{LITTLE20082828}. For threshold reactions we use the uncertainties provided by the ENDF/B-VII.1 library itself.

In order to evaluate the neutron flux spectrum in the TRIGA-RC1 irradiation facilities, we implemented the following iterative procedure:
\begin{enumerate}
\item we run the unfolding algorithm using a guess spectrum randomly selected from the set;
\item we identify, within the set of simulated spectra, the one that better fits the multi-group spectrum obtained after the unfolding;
\item we repeat the unfolding using the latter spectrum as guess, and we continue the iteration cycle starting from step 2.
\end{enumerate}
The iteration cycle ends when the spectrum that better fits the unfolding results coincides with the one used as guess. 
By applying this method to the experimental data collected at TRIGA-RC1 reactor, we find that convergence is quickly reached after one or two iterations and that the final result does not depend on the choice of the initial guess spectrum. 

The multi-group neutron flux measurement results are reported in Tabs.~\ref{Tab:LSunfolded} and \ref{Tab:CCunfolded} for Lazy Susan and Central Channel, respectively. In both cases, we achieved a precision of $\sim$2\% on total neutron flux measurement and $\lesssim$10\% on single-group neutron flux evaluation for most energy ranges (with very few exceptions up to 24\% at maximum). 

As \textit{a posteriori} check, we have recalculated the SSA of each measured reaction using the resulting multi-group fluxes, finding a very good statistical agreement with the experimental SSA values. This check highlights that the experimental SSA data supplied as input to the unfolding algorithm are self-consistent and that the multi-group flux results allow to correctly reconstruct the activation rates of a numerous set of different reactions.

In Figs.~\ref{Fig:LSmatrix} and \ref{Fig:CCmatrix} we show the correlations among flux groups. Many couples of flux groups exhibit a negative correlation, which involves a more precise knowledge of their sum than each of them taken separately. It is worth noting that the groups of fast neutrons with $E>0.5$~MeV, being constrained by the activation data from threshold reactions, are completely uncorrelated to the groups of thermal and intermediate neutrons measured through radiative capture.

In Fig.~\ref{Fig:LSunfolded}, we show the spectrum unfolding result for the Lazy Susan facility. This plot is constructed by normalizing the intra-group spectral shapes of the guess spectrum so that their subtended areas match the unfolded multi-group flux intensities. We observe that the discontinuities at the group boundaries are attributable to statistical fluctuations of the results within their uncertainty ranges. 
Since no constraint is set to force flux continuity at group edges, getting an almost continuous multi-group spectrum is a relevant achievement, that confirms the good quality of the input data and the effectiveness of the unfolding method.
In this plot, we also compare the unfolded multi-group spectrum with the guess spectrum selected at the end of the iteration cycle and renormalized to the measured total neutron flux intensity, finding a very good agreement among them. 

In Fig.~\ref{Fig:CCunfolded}, we show the same type of plot with the Central Channel results. Also in this case we obtained an almost continuous multi-group spectrum, with discontinuities at the group boundaries attributable to statistical fluctuations of the data.
As expected, the flux intensity in the Central Channel is higher and its spectrum is harder than the Lazy Susan one. In this case, we observe that the measured spectrum does not exactly match the guess one, which is slightly softer. This means that in the set of guess spectra generated with the MCNP model of the TRIGA reactor in Pavia, there isn't a spectrum with the same hardness level as that measured in the Central Channel of the TRIGA-RC1 reactor. Nevertheless, the results obtained from this analysis are self-consistent and reliable, with very bland dependence on the guess spectrum.

\section{Conclusion}
In this paper we described the experimental neutron flux characterization in the Central Channel and in the Lazy Susan irradiation facilities of the TRIGA RC-1 reactor at 100~kW power. 
The measurement of the activation rates of 41 reactions, which are neutron sensitive probes in different energy ranges, allowed for a spectrum unfolding that outputs a multi-group neutron flux measurement.

In this work, we took special care in minimizing the experimental uncertainties (both statistical and systematic). Particularly, by exploiting Monte Carlo simulations to get accurate estimates of detection efficiencies in $\gamma$-spectroscopy measurements, and by using two different HPGe detectors to reduce systematic uncertainty in absolute activity evaluation, we determined the experimental activation rates within a few percent accuracy.
The Bayesian statistical method at the basis of the unfolding algorithm has then ensured correct propagation of experimental uncertainties to the final results, which demonstrate the effectiveness of this technique in achieving high accuracy in neutron flux measurements.

Finally, the experimental measurements performed with Al-Co monitor samples highlighted that the intensity and the energy spectrum of the neutron flux inside the Lazy Susan facility is not spatially homogeneous. 
This aspect must be taken into account when evaluating the neutron dose given to a non-point size sample irradiated in this facility. 
In this respect, the development of a full-core Monte Carlo simulation of the TRIGA RC-1 reactor, to be validated with the experimental data collected in this work, will allow to accurately calculate the spectral fluence which crosses a sample interacting with the not homogeneous neutron field inside the irradiation facility.

\section*{Acknowledgments}
The authors wish to thank the ENEA TRIGA RC-1 reactor staff, together with the ENEA Radiation Protection Experts, for their support and assistance during the experimental program.
This work was carried out in the framework of the ASIF (ASI Supported Irradiation Facilities) project \url{http://www.asif.asi.it}, supported by the ``Agenzia Spaziale Italiana (ASI)'', the ``Ente per le Nuove tecnologie, l'Energia e l'Ambiente'' (ENEA), and the ``Istituto Nazionale di Fisica Nucleare'' (INFN) under ASI-ENEA ASIF implementation agreement no. 2017-22-H.0 and ASI-INFN ASIF implementation agreement no. 2017-15-H.0.
This work makes use of the \textit{Arby} software for \textsc{Geant4} based Monte Carlo simulations, that has been developed in the framework of the Milano - Bicocca R\&D activities and that is maintained by O. Cremonesi and S. Pozzi.

\section*{References}

\bibliography{Bibliography}

\end{document}